\documentclass{emulateapj}

\newcommand {\Lya}    {Ly$\alpha$}   

\newcommand {\HI}        {\ion{H}{1}}   

\newcommand {\HeII}     {\ion{He}{2}}   

\newcommand {\OVI}    {\ion{O}{6}}   
\newcommand {\OVII}   {\ion{O}{7}}
\newcommand {\OVIII}  {\ion{O}{8}}

\newcommand {\CIV}    {\ion{C}{4}}

\newcommand {\SiIV}   {\ion{Si}{4}}

\newcommand {\NeVIII}  {\ion{Ne}{8}}   
\newcommand {\NeIX}  {\ion{Ne}{9}}     

\newcommand {\HST}    {{\it HST}}

\newcommand {\etal}   {et~al.}

\begin{document}

\title{The Dispersion of Fast Radio Bursts from a \\
Structured Intergalactic Medium at Redshifts $z < 1.5$}  

\author{J. Michael Shull and Charles W. Danforth }
\affil{CASA, Department of Astrophysical \& Planetary Sciences, \\
University of Colorado, Boulder, CO 80309}

\email{michael.shull@colorado.edu, danforth@colorado.edu} 


\begin{abstract} 

We analyze the sources of free electrons that produce the large dispersion measures,
DM $\approx 300-1600$ (in units cm$^{-3}$~pc), observed toward fast radio bursts (FRBs).   
Individual galaxies typically produce DM $\sim 25-60$ cm$^{-3}$~pc from ionized  
gas in their disk, disk-halo interface, and circumgalactic medium. Toward an FRB source
at redshift $z$, a homogeneous IGM containing a fraction $f_{\rm IGM}$ of cosmological 
baryons will produce DM $= (935~{\rm cm}^{-3}~{\rm pc}) f_{\rm IGM} \, h_{70}^{-1} I(z)$, 
where $I(z) = (2/3 \Omega_m)[ \{ \Omega_m(1+z)^3 + \Omega_{\Lambda} \}^{1/2} - 1 ]$.  
A structured IGM of photoionized \Lya\ absorbers in the cosmic web produces similar 
dispersion, modeled from the observed distribution, $f_b(N,z)$, of \HI\ (\Lya-forest) 
absorbers in column density and redshift with ionization corrections and scaling relations 
from cosmological simulations.   An analytic formula for DM($z$) applied to observed 
FRB dispersions suggests that $z_{\rm FRB} \approx 0.2-1.5$ for an IGM containing a 
significant baryon fraction,  $f_{\rm IGM} = 0.6\pm0.1$.  Future surveys of the statistical 
distribution, DM($z)$, of FRBs identified with specific galaxies and redshifts can be used 
to calibrate the IGM baryon fraction and distribution of \Lya\ absorbers.  Fluctuations in 
DM at the level  $\pm10$ cm$^{-3}$~pc will arise from filaments and voids in the cosmic 
web. \\

\end{abstract} 



\section{INTRODUCTION}

Of unknown origin, but suspected to arise in explosive events at extragalactic distances,
fast radio bursts (FRBs) are radio pulses of milli-second duration that exhibit considerable 
dispersion in frequency (Lorimer \etal\ 2007; Thornton \etal\ 2013).  The observed range of
dispersion measures, DM $\equiv \int n_e \, d \ell  \approx 300-1600$ cm$^{-3}$~pc 
(Petroff \etal\ 2016)  corresponds to column densities of intervening electrons 
$N_e \approx (1-5) \times 10^{21}$~cm$^{-2}$.   Most FRBs are isolated events, with the 
exception of the repeating FRB~121102 (Spitler \etal\ 2016) which was identified with a 
low-metallicity dwarf galaxy at $z = 0.193$ (Tendulkar \etal\ 2017;  Bassa \etal\ 2017).  This
object has DM $= 557\pm2$~cm$^{-3}$~pc (Scholz \etal\ 2016), far larger than expected 
from the modeled distribution (DM $\sim 25-60$) of free electrons within the Milky Way 
(Cordes \& Lazio 2002;  Yao \etal\ 2017).  These large dispersions could arise from the 
``circumburst medium", from the host galaxy, or from the intergalactic medium (IGM).  

In this paper, we argue that the most plausible source of the high dispersions are the large 
reservoirs of ionized gas in the IGM.   The cosmological ``missing baryons problem" (Fukugita 
\etal\ 1998) has largely been solved through spectroscopic observations of diffuse baryons 
using UV absorption lines toward quasars.  These data and astrophysical analysis (Shull
\etal\ 2012) show that the low-redshift IGM still contains a substantial fraction, 
$f_{\rm IGM} \approx 0.5-0.7$, of the cosmological baryons, distributed in warm photoionized 
gas (\Lya\ forest at $T \approx 10^4$~K) and a hotter, collisionally ionized medium 
($T \approx 10^{5-7}$~K).   Thus, substantial dispersion of FRB pulses is naturally expected 
from ionized gas (H$^+$ and He$^{+2}$) that accompanies the intergalactic \HI\ (\Lya) and 
metal-line absorption systems observed toward active galactic nuclei (AGN).  

In Section 2, we assess the contributions to DM from intervening galaxies, including their 
extended gaseous halos and circumgalactic medium (CGM).  Typical galactic halos produce 
DM $\sim 25-60$ cm$^{-3}$~pc, consistent with the column density of free electrons, 
$N_e \approx 10^{20}$ cm$^{-2}$, observed in the warm ionized gas layer at the disk-halo 
interface of the Milky Way (Reynolds 1991; Gaensler \etal\ 2008).   We then compute the 
integrated column density of electrons, $N_e$, from a homogeneous IGM that contains a
substantial fraction, $f_{\rm IGM}$, of the cosmological baryons.  We derive an analytic formula 
to estimate the FRB redshift from the inferred IGM dispersion, DM$_{\rm IGM}$ and an assumed 
$f_{\rm IGM}$.  Finally, we calculate the distribution of $N_e(z)$ and DM$(z)$ from a spatially 
structured medium:  the ``cosmic web" of dark matter and baryon filaments.  Our calculation is 
based on the observed distribution of  low-redshift \HI\ and metal-line absorbers from recent 
IGM surveys (Danforth \& Shull 2008; Danforth \etal\ 2016; Shull \etal\ 2017) with the Cosmic 
Origins Spectrograph (Green \etal\ 2012) aboard the {\it Hubble Space Telescope} (\HST).     

The observed \HI\ absorbers span a wide range of column densities, $N_{\rm HI}$ (cm$^{-2}$) 
between $12.5 < \log N_{\rm HI} < 22.0$.  The nomenclature of these systems includes the Lyman 
limit systems (LLS with $\log N_{\rm HI} \geq 17.2$) and partial Lyman-limit systems (pLLS with 
$\log N_{\rm HI} \approx 16.0-17.2)$, and the much rarer ``damped \Lya\ absorbers" 
(DLAs with $\log N_{\rm HI} \geq 20.3$).  For optically thin absorbers with $\log N_{\rm HI} < 17.5$, 
large amounts of ionized hydrogen accompany the \HI.  Because the \Lya\ absorbers are highly
photoionized by the metagalactic background from AGN (Haardt \& Madau 2012), the neutral 
hydrogen absorption probes only a small fraction of the total plasma. 
In addition, the weak \Lya\ absorbers with $\log N_{\rm HI} < 13$ are quite plentiful, with an 
absorption-line frequency per unit redshift, $d{\cal N}/dz > 100$ along  sight lines toward 
background quasars.  Thus, the contribution of \Lya-forest absorbers to $N_e(z)$ is substantial
and more uniform across the sky.  

By analyzing the photoionization conditions in these \HI\ absorbers, we calculate the column 
densities of free electrons per \HI\ absorber.  We then combine the \HI\ absorber distribution
with scaling relations of baryon overdensity $(\Delta_b)$ and gas temperature ($T$) with
$N_{\rm HI}$ inferred from cosmological simulations to calculate $N_e(z)$ for a spatially 
structured IGM.  From the scaling relations, we show that \HI\ absorbers produce a flat 
distribution in integrated $N_e(z)$ over the range $12.5 \leq \log N_{\rm HI} \leq 16.0$.  
We expect a turnover and fluctuations in the DM distribution at $\log N_{\rm HI} > 15.5$, owing 
to the scarcity ($d{\cal N}/dz < 1$) of strong \HI\ absorbers.  Future FRB surveys that localize
the bursts and identify them with galaxies (with redshifts) can be used to model the IGM structure 
and to confirm the predicted large fraction of cosmological baryons in diffuse intergalactic structures.


\section{ESTIMATES OF DISPERSION MEASURES}

\subsection{Electrons in the Milky Way Halo Gas}

The observed dispersion measures toward Galactic pulsars are a standard method for
modeling the spatial distribution of electrons within the Milky Way (Cordes \& Lazio 2002).
Coupled with observations of diffuse H$\alpha$ emission, these data have led to a model
of diffuse ionized gas within a kpc of the Galactic disk.  This ionized medium at the disk-halo 
interface is often called the ``Reynolds Layer" and has a column density of free electrons 
$N_e \approx 10^{20}$~cm$^{-2}$.  From pulsar surveys, this ionized layer has been fitted 
by various surveys to an exponential density distribution, $n_e(z) = n_0 \exp(-z/h)$, with 
mid-plane densities $n_0 \approx 0.019-0.035~{\rm cm}^{-3}$ and vertical scale heights 
$h \approx 880-950$~pc (Taylor \& Cordes 1993; Cordes \& Lazio 2002).  Gaensler \etal\ (2008) 
made a joint analysis of pulsar dispersions and diffuse H$\alpha$ emission and found 
$n_0 \approx 0.014\pm0.001~{\rm cm}^{-3}$ and $h = 1830^{+120}_{-250}$~pc.   Savage \& 
Wakker (2009) obtained a revised exponential scale height  $h =1410^{+260}_{-210}$~pc.   
More recent analyses take into account spiral structure, multiple components (thin and thick 
Galactic disks, clumps, cavities).  The product of mid-plane density and scale height yields the 
``perpendicular DM" integrated from the midplane to infinity.  In the above models (Taylor \& 
Cordes 1993; Cordes \& Lazio 2002; Gaensler \etal\ 2008;  Savage \& Wakker 2009) the 
vertically integrated DM is 16.5, 33.0, 25.6 and 21.9 (cm$^{-3}$~pc) respectively.

The most recent model of the spatial distribution of Galactic free electrons (Yao \etal\ 2017)
is based on 189 pulsars with independently measured distances. The thick disk component 
has mid-plane density  $n_0 \approx 0.01132 \pm 0.00043~{\rm cm}^{-3}$ and scale height  
$h = 1673 \pm 53$~pc, corresponding to vertically integrated 
$n_0 h = 18.9 \pm 0.9$~cm$^{-3}$~pc. Their Table 14 applies their model to estimate redshifts
for 17 FRBs from inferred IGM dispersions after subtracting modeled contributions from the 
Galaxy and a standard value, DM$_{\rm host} = 100~{\rm cm}^{-3}$~pc, for the host galaxy.  
For 13 of the 17 FRBs, the modeled Galactic contribution ranges from 
$\approx 23-76$ cm$^{-3}$~pc; four sources at low Galactic latitude have higher values.

 \subsection{Electrons in Circumgalactic and Halo Gas}  
     
The best estimates of ionized gas in the low Galactic halo come from column densities $N$ of highly
ionized interstellar metal-line absorbers toward O-type stars at high Galactic latitude ($b > 40^{\circ}$)
and toward extragalactic sources such as AGN and blazars.  By fitting $N \sin b$ versus elevation above 
the disk, using UV resonant absorption lines of abundant metal ions such as \CIV, \SiIV, and \OVI, several 
groups (Sembach \& Savage 1992;   Shull \& Slavin 1994) found vertical scale heights (2-5~kpc).   
The integrated column densities through the Galactic plane allow one to estimate the total column 
density of ionized hydrogen, after correcting the metal ions for ionization fraction and metallicity.  
The largest metal-ion column densities come from X-ray absorption studies (Nicastro \etal\ 2002; 
Yao \& Wang 2005;  Wang \etal\ 2005; Fang \etal\ 2006; Anderson \& Bregman 2010) of the helium-like 
ion state of oxygen (\OVII) and its K$\alpha$ line at 21.602 \AA.  Because \OVII\ maintains a high 
ionization fraction, $f_{\rm OVII}$, over a wide range of temperatures, $5.5 < \log T < 6.3$, in collisional 
ionization equilibrium, it provides the best sensitivity to hot coronal gas.  Galactic absorption lines of \OVIII\ 
and \NeIX\ and weaker lines from ions of O, Ne, C, and N have also been detected in selected AGN sight 
lines (Nicastro \etal\ 2016; Nevalainen \etal\ 2017).  

The \OVII\ absorption lines at $z \approx 0$ observed toward many AGN are interpreted as arising from 
coronal gas in the Galactic halo (Collins \etal\ 2006; Fang \etal\ 2006; Bregman 2007).  
Typical \OVII\ column densities through the halo are $\log N_{\rm OVII} \approx 16.0-16.3$, with some 
uncertainty owing to line saturation.  For fully ionized gas with mean integrated column density 
$\langle N_{\rm OVII} \rangle = (10^{16}$~cm$^{-2}) N_{16}$ and oxygen abundance 
(O/H) $= (4.9 \times 10^{-4}) Z_{\rm O}$ relative to its solar value (O/H)$_{\odot} = 4.9\times10^{-4}$, 
we find $N_e = (2.4 \times 10^{19}~{\rm cm}^{-2}) N_{16}( f_{\rm OVII} \, Z_{\rm O})^{-1}$ corresponding 
to DM $= (7.7~{\rm cm}^{-3}~{\rm pc}) N_{16} (f_{\rm OVII} \, Z_{\rm O})^{-1}$.   With assumptions about 
the ionization and metallicity factors $(f_{\rm OVII} \, Z_{\rm O})$, the above papers estimate column 
densities of ionized hydrogen, $N_{\rm HII} \approx 10^{20}$ cm$^{-2}$ (DM $\sim 30$)
associated with the \OVII\ absorption.   
    
Strong \OVI\ absorption lines in the ultraviolet (1032 and 1038~\AA) have also been used to probe 
hot gas in the halos of external galaxies.  The COS-Halos survey (Tumlinson \etal\ 2011) of the CGM 
of galaxy halos intercepted by a sight line toward a background AGN found mean \OVI\ column 
densities $\langle N_{\rm OVI} \rangle \approx 10^{14.5}$~cm$^{-2}$ for actively star-forming galaxies.
Employing corrections for metallicity (O/H) and  \OVI\ ionization fraction, $f_{\rm OVI} \approx 0.2$ in 
collisional ionization equilibrium, we estimate a column density of ionized gas (and electrons) of
$N_e \approx (3.2 \times 10^{18}~{\rm cm}^{-2}) Z_{\rm O}^{-1}$.  At the metallicities ($Z_{\rm O} = 0.1-0.5$) 
observed in Galactic high-latitude high-velocity clouds (Wakker \etal\ 1999; Shull \etal\ 2009; Fox \etal\ 2014)
the disk-halo interface gas traced by \OVI\ produces dispersions DM $\sim 5-10~{\rm cm}^{-3}$~pc.    
The frequencies per unit redshift of galaxy halos, LLS, DLAs, and other strong absorbers are too low 
to explain the large FRB dispersions.   For example, the incidence of LLS at $z \leq 0.5$ has been 
measured by \HST/COS (Shull \etal\ 2017) to be $(d{\cal N}/dz)_{\rm LLS} = 0.36^{+0.20}_{-0.13}$. 
If each LLS contributed $N_e \approx 10^{17.5}$~cm$^{-2}$, the mean integrated dispersion would be 
$\langle {\rm DM} \rangle \approx 0.04$ cm$^{-3}$~pc toward FRBs at $z \approx 1$.  Similarly, if intervening 
galactic halos and their CGM each produce DM $\sim 30-50~{\rm cm}^{-3}$~pc, an FRB sight line would 
need to intercept 10-30 such galaxies to accumulate sufficient electron column densities sufficient  to explain 
the large FRB dispersions, DM $= 300-1600~{\rm cm}^{-3}$~pc.   For all these reasons, we now explore 
dispersion in the the baryon-rich IGM.
    
\subsection{Electrons in a Homogeneous IGM}

As a first estimate of the IGM dispersion, we compute the integrated column density of electrons 
in a homogeneous IGM, whose mass density increases as $(1+z)^3$ with redshift.  
The mean cosmological baryon density has been well constrained by measurements of the 
primordial D/H ratio (Cooke \etal\ 2016) and the acoustic spectrum  of the cosmic microwave 
background (Planck Collaboration 2016).  The cosmological baryon density is independent of the 
Hubble constant $H_0 = (100~{\rm km~s}^{-1}~{\rm Mpc}^{-1})h$ and its scaling factor ($h$),
$\bar{\rho}_b = \Omega_b \rho_{\rm cr} = 4.17 \times 10^{-31}~{\rm g~cm}^{-3}$, based on
$\Omega_b h^2 = 0.02217$ and a critical (closure) density 
$\rho_{\rm cr} = (1.879 \times 10^{-29}~{\rm g~cm}^{-3})h^2$.  With a primordial helium abundance 
$Y \approx 0.2449$ by mass (Aver \etal\ 2013), this mass density corresponds to a number density 
of hydrogen,
\begin{equation}
   \bar{n}_H = \frac { \rho_b (1-Y)}{m_H} \approx (1.88 \times 10^{-7}~{\rm cm}^{-3}) f_{\rm IGM} (1+z)^3 \; , 
\end{equation}
where $f_{\rm IGM} \approx 0.5-0.7$ is the likely fraction of diffuse baryons in the IGM (Shull \etal\ 2012). 
The number density of electrons is $\bar{n}_e = 1.167 \bar{n}_H$, accounting for H$^+$ and He$^{+2}$, 

To find the column density, $N_e$, of electrons toward a source at redshift $z$, we use the relationship 
between proper length and redshift in a flat $\Lambda$CDM cosmology,
\begin{equation}
    \frac {d \ell} {dz} = \frac {c}{H_0} (1+z)^{-1} \left[ \Omega_m (1+z)^3 + \Omega_{\Lambda} \right]^{-1/2}  \; .
\end{equation}
Here, we adopt $\Omega_m \approx 0.3$ and $\Omega_{\Lambda} \approx 0.7$ as the fractional contributions 
of matter and dark energy to closure density and $H_0 = (70~{\rm km~s}^{-1}~{\rm Mpc}^{-1})h_{70}$.  
The integrated column density of electrons out to redshift $z$ is then
\begin{eqnarray}
    N_e (z) &=& (1.167)(1.87 \times 10^{-7}~{\rm cm}^{-3})  f_{\rm IGM}  \frac {c}{H_0} \nonumber \\
          & \times & \int_0^z \frac {(1+z)^2 \, dz} { \left[ \Omega_m (1+z)^3 + \Omega_{\Lambda} \right]^{1/2} }
            \nonumber \\
       &=& (2.88 \times 10^{21}~{\rm cm}^{-2}) f_{\rm IGM} \,  h_{70}^{-1} \, I(z)  \; .
 \end{eqnarray}
 The redshift integral can be done analytically, 
 \begin{eqnarray}
  I(z)  &=& \int_0^z \frac { (1+z)^2 \, dz} { \left[ \Omega_m (1+z)^3 + \Omega_{\Lambda} \right]^{1/2} }  \nonumber \\
       &=& \frac {2}{3 \Omega_m} \left[ \{ \Omega_m (1+z)^3 + \Omega_{\Lambda} \} ^{1/2} - 1 \right]   \; ,
 \end{eqnarray}
with values of 0.686, 1.69, and 2.94 at $z = 0.5, 1.0, 1.5$ respectively.  For a homogeneous IGM, the 
dispersion measure can be expressed as
\begin{equation}
    {\rm DM} = (935~{\rm cm}^{-3}~{\rm pc}) f_{\rm IGM} \, h_{70}^{-1} I(z)  \; .
\end{equation}
We invert this relation for $I(z)$ to find an analytic formula for the FRB redshift based on the 
inferred dispersion, DM$_{\rm IGM}$, contributed by the IGM,
\begin{equation}
    (1+z) = \left[ \frac { \{ (3 \Omega_m d /2) + 1 \}^2 - \Omega_{\Lambda}} { \Omega_m} \right]^{1/3} \; .  
\end{equation}  
Here, we  have defined the dimensionless dispersion parameter 
\begin{equation}
    d = \left[   \frac { {\rm DM}_{\rm IGM} \, h_{70} } { (935~{\rm cm}^{-3}~{\rm pc}) f_{\rm IGM}} \right]  \; \; .
\end{equation} 
Table 1 gives our estimated redshifts for these 17 FRBs and 8 others from recent papers 
(Bhandari \etal\ 2017; Bannister \etal\ 2017; Caleb \etal\ 2017).  We use equations (6) and (7) 
with values of IGM baryon fraction, $f_{\rm IGM} = 0.6 \pm 0.1$, consistent with \HST/COS baryon 
surveys and cosmological simulations (Shull \etal\ 2012).   For the IGM dispersion, we follow the 
methodology in Yao \etal\ (2017) in which
DM$_{\rm IGM} = {\rm DM}_{\rm obs} - {\rm DM}_{\rm Gal} - {\rm DM}_{\rm Host}$.  We use their 
model of the electron distribution within the Milky Way for the Galactic contribution, DM$_{\rm Gal}$,
and we adopt a constant dispersion DM$_{\rm Host} = 100$ cm$^{-3}$~pc  (their assumption) 
for the FRB host galaxy.  This assumption is uncertain, as DM$_{\rm Host}$ is larger than most 
values within the Milky Way.  
For 17 FRBs in their study, Yao \etal\ (2017) inferred redshifts ranging from $z = 0.238$ to $z = 2.059$
with median $z = 0.687$.   The median modeled IGM dispersion is 490~cm$^{-3}$~pc 
(range 170 to 1469~cm$^{-3}$~pc) and the median observed dispersion is 776~cm$^{-3}$~pc 
(range 375 to 1629~cm$^{-3}$~pc).   In general, our method produces lower redshifts for the 17 FRBs, 
ranging from  $z = 0.254^{+0.043}_{-0.032}$ for the burst with the lowest dispersion 
(DM$_{\rm IGM} \approx 170$ cm$^{-3}$~pc) to $z = 1.38^{+0.196}_{-0.147}$ for the largest dispersion
(DM$_{\rm IGM} \approx 1469$~cm$^{-3}$~pc).  The error bars on these redshifts are propagated  
from the assumed uncertainty in IGM baryon fraction.  They do not include errors in the cosmological
parameters  ($\Omega_m$,  $\Omega_{\Lambda}$, $H_0$) or assumptions about the subtracted
portions of DM from the Galaxy or FRB host galaxy.  

\subsection{Electrons Associated with the \Lya\ Forest}
  
Next, we compute the electrons associated with a more realistic, structured IGM based on the 
distribution of \Lya\ absorbers measured in recent surveys.  Figure 1 shows examples of low-redshift \HI\ 
absorbers in the spectra of four blazars.  From a recent survey with \HST/COS (Danforth \etal\ 2016),
the typical frequency per unit redshift of \HI\ absorbers is $d{\cal N}/dz \approx 50-150$ for weak systems 
in the \Lya\ forest ($\log N_{\rm HI} \approx 12.5-14.0)$.   A survey of higher column density absorbers 
(Shull \etal\ 2017) characterized the distribution of low-redshift LLS ($\log N_{\rm HI} \geq 17.2$) and 
pLLS ($\log N_{\rm HI} \approx 16.0-17.2)$.  These absorbers are considerably rarer, with 
$(d{\cal N}/dz)_{\rm LLS} \approx 0.36^{+0.20}_{-0.13}$ and 
$(d{\cal N}/dz)_{\rm pLLS} \approx 1.69 \pm 0.23$.


\begin{figure*}
\includegraphics[angle=0,scale=0.75] {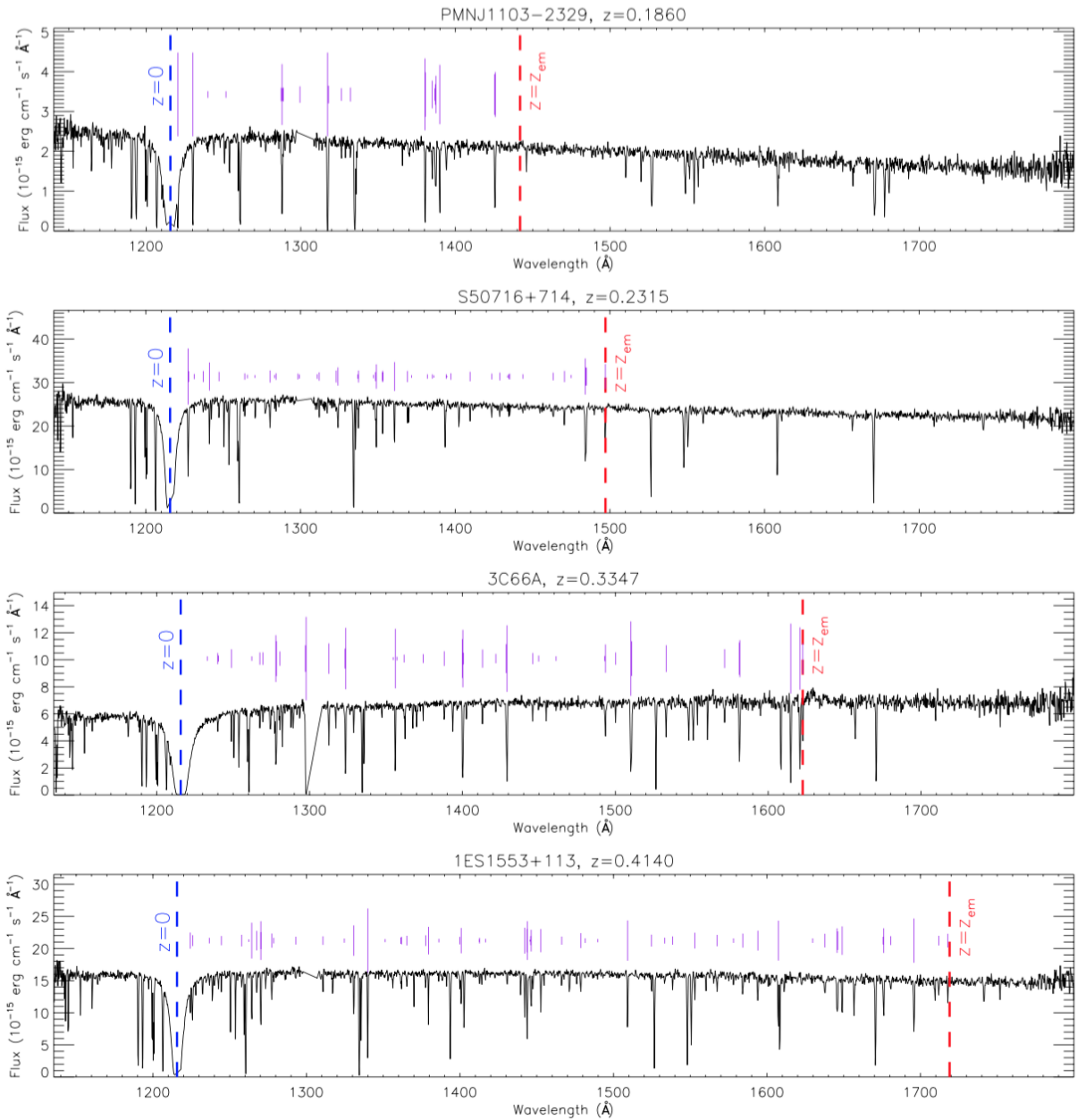} 
\caption{\small{ HST/COS spectra of the low-redshift \HI\ (\Lya) absorbers toward four blazars at 
$z = 0.1860$ (PMN\,J1103-2329), $z = 0.2315$ (S5\,0716+714),  $z = 0.3347$ (3C\,66A),
and $z \approx 0.414$ (1ES\,1553+113).  The redshift of PG~1553+113 (Danforth \etal\ 2010) 
was inferred statistically from the wavelength at which no further \Lya\ lines are seen. Tick marks 
above the continuum indicate intervening \Lya\ absorbers, with length denoting the \HI\ column density. }
 }
\end{figure*}


To compute the ionized gas (and electrons) associated with these \Lya\ absorbers, we employ scaling 
relations found in simulations of the low-redshift IGM.   These formulae allow us to relate the baryon 
overdensity of an absorber, $\Delta_b = \rho_b / \bar{\rho}_b$, to \HI\ column density 
$N_{\rm HI} = (10^{14}~{\rm cm}^{-3})N_{14}$ and gas temperature $T$.  The relations are expressed
as power laws, 
\begin{equation}
   \Delta_b = (36.9) N_{14}^{\alpha}  \; \; \; {\rm and} \; \; \;  T = (5000~{\rm K}) \Delta_b^{\beta} \; \; .
\end{equation}
In our IGM simulations (Shull \etal\ 2015) we found $\alpha \approx 0.65$ and $\beta \approx 0.60$.  
This temperature relation has been confirmed by other groups, although several simulations found 
slightly larger values of $\alpha = 0.741\pm0.003$ (Dav\'e \etal\ 2010),  $0.786\pm0.001$ 
(Tepper-Garcia \etal\ 2012), and $0.770\pm0.022$ (Gaikwad \etal\ 2017).  Baryon overdensity is 
defined relative to the mean (co-moving) cosmological baryon density, 
$\bar{\rho}_b = \Omega_b \, \rho_{\rm cr} = 4.17 \times10^{-31}$~g~cm$^{-3}$.  
For an \HI\ absorber at redshift $z$ and overdensity $\Delta_b$, the hydrogen number density is
 \begin{equation}
      n_H (z) = \frac { \rho_b (1-Y)} {m_H} (1+z)^3  \approx 
             (1.88 \times 10^{-7}~{\rm cm}^{-3})  \Delta_b (1+z)^3  \; \; .
\end{equation} 
We assume that the plasma in this absorber is in photoionization equilibrium with the metagalactic 
ionizing background, with hydrogen photoionization rate $\Gamma_H$ (s$^{-1}$) and the low-density 
(case-A) hydrogen recombination rate coefficient $\alpha_H^{\rm (A)}$(cm$^3$~s$^{-1}$).  In the 
approximation $x_{\rm HI} \ll 1$, the hydrogen neutral fraction is given by
\begin{equation}
   x_{\rm HI} = \frac {n_{\rm HI}} {n_H} \approx \frac {n_e \alpha_H^{(\rm A)} } {\Gamma_H } \; . 
\end{equation}
We approximate $\Gamma_H \approx (4.6 \times 10^{-14}~{\rm s}^{-1})(1+z)^{4.4}$ for the 
hydrogen photoionization rate over the range $0 < z < 0.47$ (Shull \etal\ 2015) and adopt 
$\alpha_H^{\rm (A)} \approx (4.09 \times 10^{-13}~{\rm cm}^3~{\rm s}^{-1}) T_4^{-0.726}$ 
at temperature $T = (10^4~{\rm K})T_4$.   From $T(\Delta_b)$ and $\Delta_b (N_{14})$ 
in equation (8), we find
\begin{eqnarray}
   x_{\rm HI}  & \approx & (3 \times 10^{-6}) \Delta_b^{0.564} (1+z)^{-1.4} \nonumber \\
       &  \approx & (2.29 \times 10^{-5})  N_{14}^{0.37} (1+z)^{-1.4}  \; . 
 \end{eqnarray}
The total column density of ionized hydrogen accompanying the \HI\ is
\begin{equation}
    N_{\rm H} \approx \frac {N_{\rm HI}} {x_{\rm HI} } \approx (4.4 \times 10^{18}~{\rm cm}^{-2}) 
          N_{14}^{0.63} (1+z)^{1.4}  \;  , 
\end{equation}
with a characteristic absorber size 
\begin{equation} 
   L \approx \frac { N_{\rm HI} } {n_H \, x_{\rm HI}} \approx (200~{\rm kpc}) 
         N_{14}^{-0.02} (1+z)^{-1.6}  \; ,
 \end{equation}
nearly constant with \HI\ column density.  These scaling relations suggest that, at fixed redshift,
higher column density absorbers have larger neutral fractions. The higher redshift absorbers 
are smaller and more ionized.


\begin{figure*}
\includegraphics[angle=0,scale=0.65] {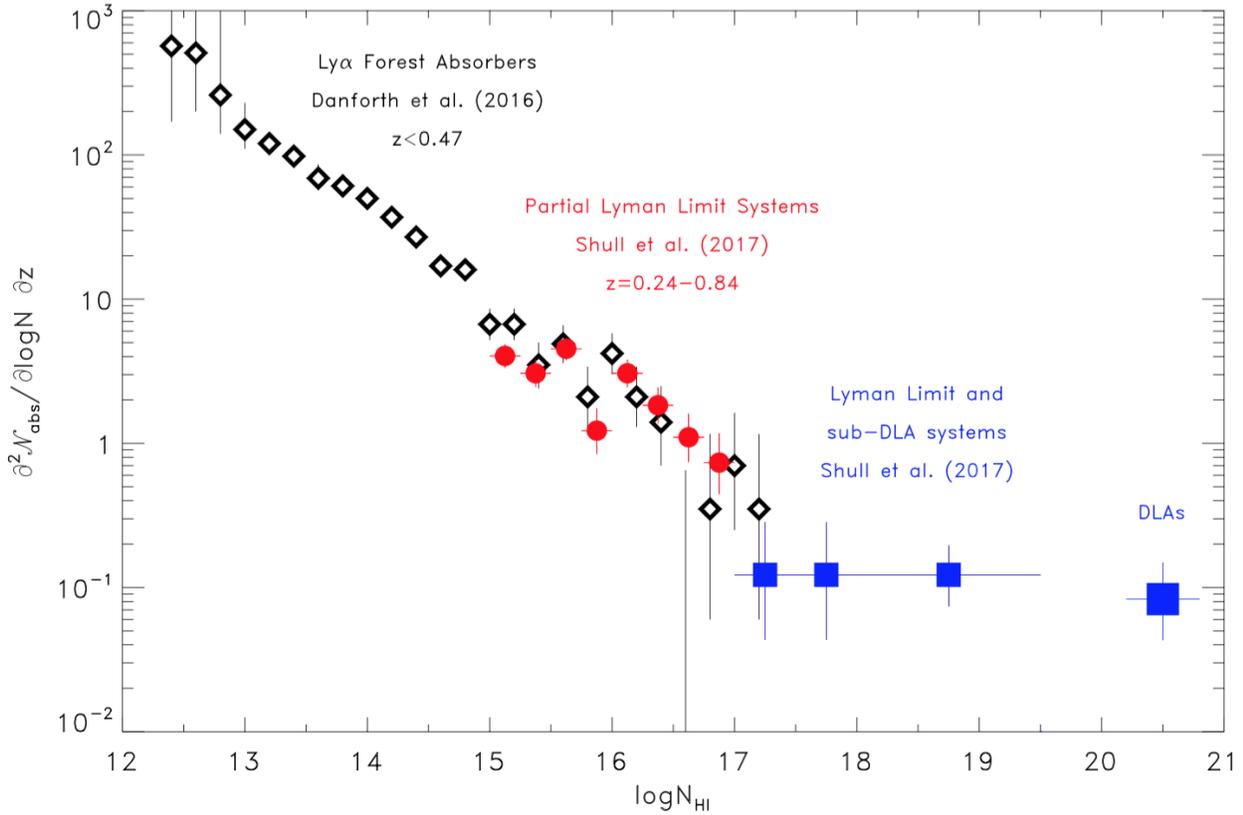} 
\caption{\small{ Distribution of low-redshift \HI\ absorbers per unit redshift and column density ($\log N_{\rm HI}$).
Solid diamonds (black) are from the \HST/COS survey (Danforth \etal\ 2016) of \Lya\ forest at $0 < z < 0.47$.
Filled circles (red) are results from the \HST/COS survey (Shull \etal\ 2017) of strong \HI\ absorbers at 
$0.24 < z < 0.84$ (pLLS) and $0.24 < z < 0.48$ (LLS).   The four solid squares (blue) show values for
sub-DLA and DLA systems ($0 < z < 0.48$).  The distribution of \Lya-forest absorbers is well fitted with a 
differential distribution, $f(N,z) \propto N^{-\beta}$, with $\beta  = 1.65\pm0.02$. 
\vspace{1cm}  } 
 }
\end{figure*}


We  now integrate the electron column densities over the distribution of \HI\ absorbers found in 
the recent \HST/COS surveys (Danforth \etal\ 2016;  Shull \etal\ 2017).  As shown in Figure 2,
a power-law fit to the differential distribution in \HI\ column density per unit redshift is
$f(N,z) \equiv (d ^2 {\cal N}  / {d N_{\rm HI} \, dz}) \approx 50 N_{14}^{-1.65} (1+z)^{\gamma}$.
We adopt a redshift-evolution factor, $\gamma \approx 1.24$, from a new fit to the absorbers
with $\log N_{\rm HI} \geq 15$ (Danforth \etal\ 2016), and multiply by a factor of 1.167 for the
electrons donated from He$^{+2}$.  After the \HeII\ reionization epoch at  $z \approx 2.7-3.2$,
\Lya\ forest clouds have far more He$^{+2}$ than He$^+$ (Shull \etal\ 2010).  We arrive at the 
electron column density, $N_e(z)$, out to redshift $z$, integrated over column densities from 
$N_1 = 10^{12.5}$ to $N_2 = 10^{16.0}$ cm$^{-2}$,
\begin{eqnarray}
    N_e(z) &=& (1.083)(4.4 \times 10^{18}~{\rm cm}^{-2}) (50) \nonumber \\
        & \times &  \int_0^z (1+z)^{\gamma} \, dz  \int_{N_1}^{N_2} N_{14}^{-0.02} \, dN_{14} \nonumber  \\
     &=& (1.65 \times 10^{21}~{\rm cm}^{-2})  \frac  {\left[ (1+z)^{\gamma+1} - 1 \right] } {(\gamma+1)} \; .
\end{eqnarray} 
This column density translates to 
DM $= (534~{\rm cm}^{-3}~{\rm pc}) (\Delta z)_{\rm eff}$, where the effective redshift accounts for
the $(1+z)^{\gamma}$ evolution of the \Lya\ absorber frequency with redshift.  For $\gamma = 1.24$, 
we have $(\Delta z)_{\rm eff} = 0.66$ out to $z = 0.5$ and 1.66 out to $z = 1.0$.   Additional baryons 
reside in the WHIM probed by \OVI\ and Broad \Lya\ Absorbers.

  
 \section{SUMMARY AND DISCUSSION}
 
 Many physical locations have been suggested to explain the large pulse dispersions observed toward 
 FRBs.   In the most recent DM model for the Galactic electron distribution (Yao \etal\ 2017) the 
 Milky Way typically contributes DM $\sim 25-75$ cm$^{-3}$~pc, consistent with other measurements of 
 ionized gas.  Studies of metal ions at the disk-halo interface and Galactic halo find electron column 
 densities $N_e \approx 10^{20}~{\rm cm}^{-3}$ (DM $\sim 30$).  Thus, for the observed FRB range, 
 DM $\approx 300-1600$ cm$^{-3}$~pc, the host galaxy and the IGM are probably the dominant 
 contributors to the dispersion.   Yao \etal\ (2017) adopted a standard contribution, 
 DM$_{\rm Host} = 100$ cm$^{-3}$~pc, from the host galaxy.  After subtracting the modeled
Galactic DM, they attribute the residual DM to plasma in the IGM.  Their values for DM$_{\rm IGM}$ are 
in good agreement with values derived in Sections 2.3--2.4.  It would be difficult for FRBs at cosmological 
distances to avoid having large dispersions, given the substantial baryon fractions inferred to reside in
the IGM (Shull \etal\ 2012).   For example, the \Lya\ forest likely contains 30\% of the cosmic baryons in 
a warm ($10^4$~K) photoionized phase. Absorption from  high ions such as \OVI\ and \NeVIII\ (and broad 
\Lya\ absorbers) suggest a similar contribution from hotter gas ($10^5 - 10^7$~K) in a phase called the 
WHIM (warm-hot intergalactic medium).  

The agreement between estimates of cosmological dispersion measures, DM($z$), and baryon 
measurements from UV spectroscopic surveys, provides additional evidence that the IGM contains 
a substantial fraction, $f_{\rm IGM} > 0.5$, of diffuse baryons.  However, both are model-dependent 
estimates.  One important prediction of our calculation is the flat contribution to DM$(z)$ from \Lya-forest
absorbers across a wide range of column densities, $12.5 <  \log N_{\rm HI} < 16$.  Further progress 
in using FRBs as IGM probes will require identifying their host galaxies and obtaining redshifts.  One 
can then assemble a statistical sample to relate DM$(z)$ to electrons and \HI\ absorbers in the IGM.  
Fluctuations in DM from a structured IGM are expected at the level of $\pm10$ cm$^{-3}$~pc from 
the (Mpc-scale) filaments and (10-30~Mpc) voids produced by gravitational instability 
(Dav\'e \& Oppenheimer 2010;  Smith \etal\ 2011).   We also expect to see fluctuations at the level of 
DM $\approx 25$~cm$^{-3}$~pc, owing to the scarcity of \HI\ absorbers with $\log N > 15$.  Our
recent  \HST/COS survey (Shull \etal\ 2017) of strong \Lya\ absorbers ($0.24 < z < 0.48$) found line 
frequencies $(d{\cal N}/dz) \approx 1.69 \pm 0.23$ at $\log N_{\rm HI} \geq 16.0$ and
$(d{\cal N}/dz) \approx 4.95 \pm 0.39$ at $\log N_{\rm HI} \geq 15.0$.  

\vspace{0.3cm}

\noindent
We now summarize the main results of our survey: 
\begin{enumerate}     

\item The large observed dispersions, DM $\approx 300-1600~{\rm cm}^{-3}~{\rm pc}$, of FRB 
pulses are unlikely to arise in the ionized gaseous layers of galaxies or in their halos, but the 
integrated dispersion of electrons in the IGM naturally produces such values.  For a homogeneous
IGM containing a fraction $f_{\rm IGM}$ of cosmological baryons, the average dispersion measure
DM $\approx (935~{\rm cm}^{-3}~{\rm pc}) f_{\rm IGM} h_{70}^{-1} I(z)$, where 
$I(z) = (2/3 \Omega_m)[ \{ \Omega_m(1+z)^3 + \Omega_{\Lambda} \}^{1/2} - 1 ]$.  

\item Our model for DM$_{\rm IGM}(z)$ provides a convenient analytic formula for estimating FRB 
redshifts,  $(1+z) = \Omega_m^{-1/3} \left[  \{ (3 \Omega_m d / 2) + 1 \}^2 - \Omega_{\Lambda} \right]^{1/3}$.
The dimensionless  parameter 
$d = \left[  {\rm DM}_{\rm IGM}  \,  h_{70}/ (935~{\rm cm}^{-3}~{\rm pc}) f_{\rm IGM} \right]$ depends on the 
dispersion attributed to the IGM, after subtracting contributions from the Milky Way and FRB host galaxy.  

\item We also analyzed a more sophisticated IGM model using UV spectroscopic observations of the 
bivariate distribution $(d^2 {\cal N} / dz \, dN_{\rm HI})$ of intergalactic \HI\ absorbers in column density 
and redshift,  together with scaling relations from cosmological simulations of  \HI\ column densities 
with baryon overdensity, temperature, and ionization state.  This model finds 
DM $\approx (534~{\rm cm}^{-3}~{\rm pc}) (\Delta z)_{\rm eff}$ over effective pathlength
$(\Delta z)_{\rm eff} = 0.66$ integrated out to $z = 0.5$ and 1.66 to $z = 1.0$, consistent with
$f_{\rm IGM} \approx 0.6$.  

\item Our model of a structured IGM, together with scaling relations, predicts nearly equal 
contributions to the integrated DMs across a wide range of \HI\ column densities, 
$12.5 \leq \log N_{\rm} < 16$.   One expects fluctuations at the level of 
DM $\approx 25~{\rm cm}^{-3}~{\rm pc}$ from absorbers $\log N_{\rm HI} > 15.5$, 
when the number of strong absorbers per unit redshift drops below $d {\cal N} / dz  < 1$.   
 
 \item With a sufficiently large sample of FRBs identified with galaxies (and redshifts) one can 
calibrate the fraction of baryons in the diffuse IGM.  Strong filaments and voids in 
the cosmic web would appear as weak DM fluctuations at the level of 10~cm$^{-3}$~pc.  
   
\end{enumerate}

\acknowledgments
  
\noindent
The IGM data originated from individual and survey observations of AGN taken with the Cosmic 
Origins Spectrograph on the {\it Hubble Space Telescope}.  We appreciate helpful discussions with
Jeremy Darling, Shri Kulkarni, and John Stocke.  In early stages, this research was supported by 
grants HST-GO-13301.01.A and HST-GO-13302.01.A from the Space Telescope Science Institute 
to the University of Colorado  Boulder.  More recent work was carried out through academic support
from the University of Colorado.


 \clearpage


\begin{deluxetable}{lclcccc}
\tabletypesize{\scriptsize}
\tablecaption{\bf Redshift Estimates from FRB Dispersions\tablenotemark{a}  }  
\tablecolumns{7}
\tablewidth{0pt}
\tablehead{ \colhead{No.}  &  \colhead{FRB} & \colhead{DM$_{\rm obs}$} & \colhead{DM$_{\rm Gal}$} 
  & \colhead{DM$_{\rm IGM}$}  & \colhead{$z_{\rm Yao}$} & \colhead{$z_{\rm SD}$}   \\
  &  & (cm$^{-3}$~pc)  &  (cm$^{-3}$~pc) & (cm$^{-3}$~pc) & (Model) & (Model)   }  
\startdata
1   & FRB\,121102  & $557\pm2$   &   287   & 170     &  0.238  &  $0.254^{+0.043}_{-0.032}$    \\
2   & FRB\,010724  & $375\pm3$   & 32.7    &  181    &  0.254  &  $0.268^{+0.044}_{-0.033}$    \\
3   & FRB\,130628  & $470\pm1$   & 47.0    &  323    &  0.453  &  $0.434^{+0.069}_{-0.052}$    \\
4   & FRB\,010621  & $745\pm10$ &  321.6 & 323    &  0.453  &  $0.435^{+0.069}_{-0.052}$    \\
5   & FRB\,150418  & $776\pm5$   &  325.5 &  351    &  0.492  &  $0.437^{+0.069}_{-0.052}$    \\
6   & FRB\,120127  & $553\pm3$   &  20.6   & 432     &  0.607  &  $0.549^{+0.084}_{-0.063}$     \\
7   & FRB\,140514  & $563\pm6$   &  24.2   &  438    &  0.615  &  $0.555^{+0.085}_{-0.064}$     \\
8   & FRB\,131014  & $779\pm1$   &  220.2 &  459    &  0.643  &  $0.575^{+0.088}_{-0.066}$     \\
9   & FRB\,110523  & $623\pm6$   &  33.0   & 490     &  0.687  &  $0.605^{+0.092}_{-0.069}$     \\
10 & FRB\,110627  & $723\pm3$   & 33.6    &  589    &  0.826  &  $0.698^{+0.104}_{-0.079}$     \\
11 & FRB\,101125  & $790\pm3$   &   75.9  & 614     &  0.861  &  $0.721^{+0.107}_{-0.081}$      \\
12 & FRB\,130729  & $861\pm2$   &  25.4   & 736     &  1.031  &  $0.827^{+0.121}_{-0.091}$      \\
13 & FRB\,090625  & $900\pm1$   &  25.5   & 774     &  1.085  &  $0.859^{+0.126}_{-0.095}$      \\
14 & FRB\,130626  & $952\pm1$   &  65.1   &  787    &  1.104  &  $0.870^{+0.127}_{-0.096}$      \\
15 & FRB\,110220  & $944\pm5$   &  24.1   & 820     &  1.150  &  $0.897^{+0.131}_{-0.098}$       \\
16 & FRB\,110703  & $1104\pm7$ &  23.1   & 981     &  1.375  & $1.025^{+0.148}_{-0.111}$       \\
17 & FRB\,121002  & $1629\pm2$ & 60.5    &1469    &  2.059  & $1.380^{+0.196}_{-0.147}$      \\
     &    &    &    &    &    &       \\
18 & FRB\,150610  & $1593.9\pm0.6$  & 122  &  1372  &  1.2    & $1.313^{+0.187}_{-0.140}$     \\
19 & FRB\,151206  & $1909.8\pm0.6$  &   16   &  1650  &  1.5    & $1.502^{+0.212}_{-0.160}$     \\
20 & FRB\,151230  & $960.4\pm0.5$    &    38  &    822  &  0.8    & $0.899^{+0.131}_{-0.099}$     \\
21 & FRB\,160102  & $2596 .1\pm0.3$ &   13   &  2483  &   2.1   & $2.018^{+0.283}_{-0.212}$      \\
22 & FRB\,170107  & $609.5\pm5$        &   27   &    483  &  \dots & $0.598^{+0.091}_{-0.069}$      \\
23 & FRB\,160317  & $1165\pm11$       & 395  &     670  &  0.7    & $0.770^{+0.114}_{-0.086}$      \\
24 & FRB\,160410  & $278\pm3$            & 62.5 &    115   &  0.2    & $0.180^{+0.032}_{-0.023}$      \\
25 & FRB\,160608  & $682\pm7$            & 310  &    272   &  0.4    & $0.377^{+0.066}_{-0.045}$      \\
   
\enddata  

\tablenotetext{a} {Observed and IGM dispersion measures, DM$_{\rm obs}$ and 
DM$_{\rm IGM}$, for 17 FRBs (Petroff \etal\ 2016; Yao \etal\ 2017) listed in order 
(\#1 - \#17) of increasing IGM dispersion and estimated redshift.  The next 8 FRBs 
are taken from recent papers:  \#18 - \#21 (Bhandari \etal\ 2017);  
\#22 (Bannister \etal\ 2017);  \#23 - \#25 (Caleb \etal\ 2017).  
Column 5 gives the IGM dispersion, after subtracting contributions from the Milky 
Way Galaxy (column 4) and FRB host galaxy (constant 100~cm$^{-3}$~pc).  
Column (6) gives estimated redshift quoted by Yao \etal\ (2017) and other groups.
Column (7) gives our estimate of the redshift using equations [6] and [7] with
$h_{70} = 1$ and baryon fraction $f_{\rm IGM} = 0.6 \pm 0.1$.    Errors on redshift
do not include systematic effects from cosmological model (about 1\%) or DM 
subtractions for Milky Way and FRB host galaxy (see discussion in Section 2.3).  }

\end{deluxetable}


\end{document}